\documentclass[%
aps,amsmath,amssymb,
reprint,%
]{revtex4-1}

\usepackage[francais]{babel}
\usepackage{graphicx}
\usepackage{dcolumn}
\usepackage{bm}
\usepackage{caption}
\usepackage[font=small,labelfont=bf,justification=justified]{caption}

\usepackage[utf8]{inputenc}
\usepackage[T1]{fontenc}
\usepackage{mathptmx}
\usepackage{color}
\usepackage[francais]{babel}
\usepackage{lmodern}
\usepackage{amsmath}
\usepackage{amsfonts}
\usepackage{amsthm}
\usepackage{caption}
\usepackage{subcaption}
\usepackage{multirow}

\begin{document}
	\title{Accessing low-energy magnetic microstates in symmetry-broken isolated square artificial spin ice vertices with magnetic field}

\author{Neeti Keswani}
\affiliation{Department of Physics, Indian Institute of Technology, Delhi, New Delhi 110016, India}
\author{Ranveer Singh}
\affiliation{Institute of Physics, Sachivalaya Marg, Bhubaneswar 751005, Odisha, India}
\author{Yoshikata Nakajima}
\affiliation{Bio-Nano Electronics Research Centre, Toyo University, Saitama 3508585, Japan}
\author{Sakthi Kumar}
\affiliation{Bio-Nano Electronics Research Centre, Toyo University, Saitama 3508585, Japan}
\author{Tapobrata Som}
\affiliation{Institute of Physics, Sachivalaya Marg, Bhubaneswar 751005, Odisha, India}
\author{Pintu Das}
\email{pintu@physics.iitd.ac.in}
\affiliation{Department of Physics, Indian Institute of Technology, Delhi, New Delhi 110016, India}

  	\email{pintu@physics.iitd.ac.in} 
	\date{\today}
	
	\begin{abstract}

In artificial spin ice systems, an interplay of defects and dipolar interactions is expected to play important roles in stabilizing different collective magnetic states. In this work, we investigated the magnetization reversal of individual defective square artificial spin ice vertices where defects break four-fold rotational symmetry of the system. By varying the angle between the applied field and the geometrical axis of the vertices, we observe a change in energy landscape of the system resulting into the stabilization of collective low-energy magnetic states. We also observe that by changing the angle, it is possible to access different vertex configurations. Micromagnetic simulations are performed for varying angle as well as external field, the results of which are consistent with the experimental data. 
	\end{abstract}
	\maketitle
	\section{\label{sec:level1}INTRODUCTION}
\begin{figure*}
	\includegraphics[width=0.8\textwidth]{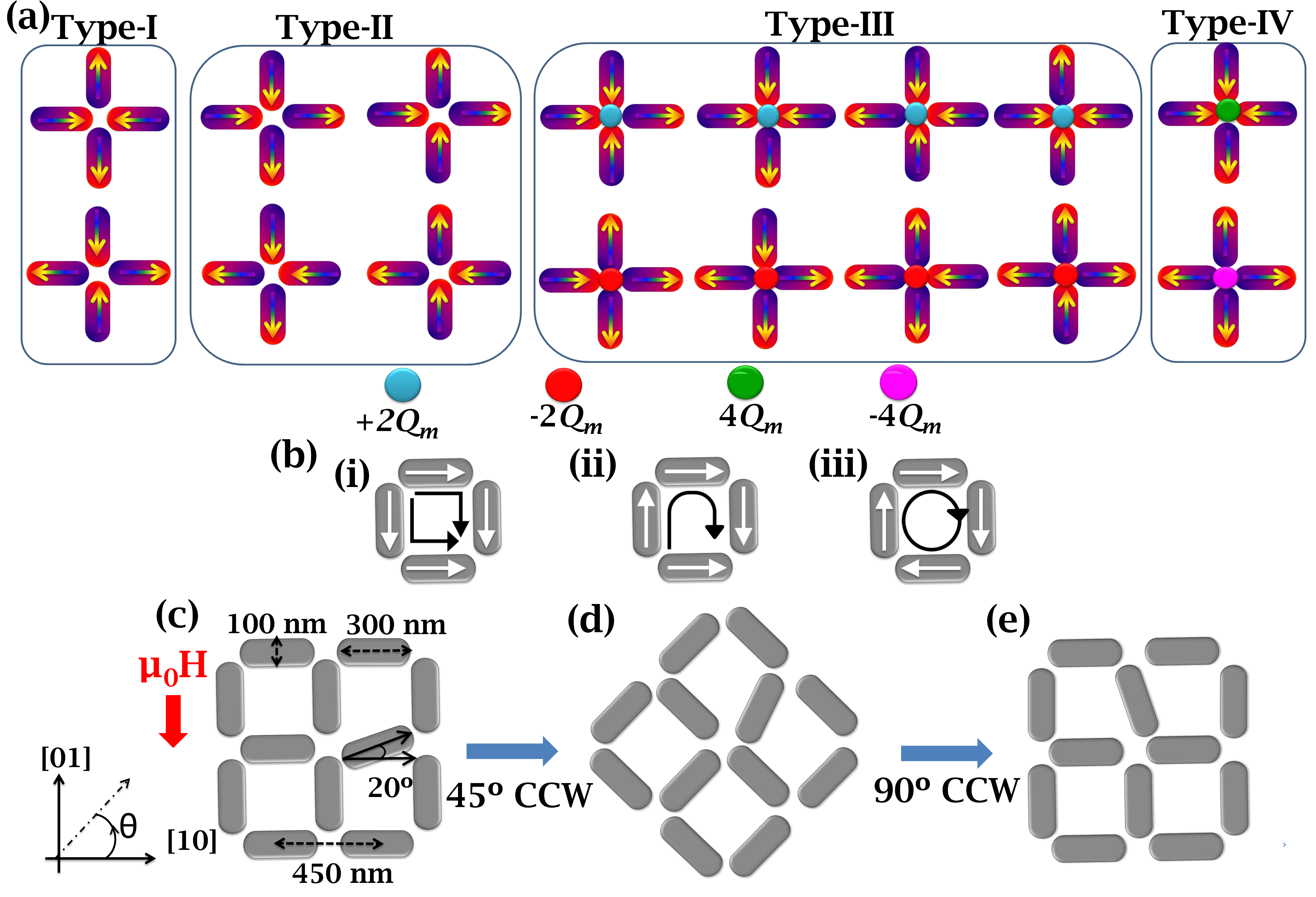}
    \caption{\label{sem} (a) Illustration of 16 possible spin states at a vertex of two-dimensional square artificial spin ice system. Degenerate states are grouped into four different types of increasingly higher energies. The different colored circles at the vertex for type-III states represent the different net magnetic charges at the vertex. (b) Schematic of (i) onion state (ii) horse-shoe state, and (iii) microvortex states~\cite{keswani2018magnetization,keswani2019micromagnetic}. (c)-(e) Schematics of rotation of sample with fixed magnetic field direction.}
	\end{figure*}

Artificial spin-ice (ASI) systems are lithographically patterned arrangements of interacting magnetic nanostructures that were introduced for investigating the effects of geometric frustration in a controlled manner~\cite{wang2006artificial,nisoli2013colloquium,rougemaille2019cooperative}. These are 2D arrays of nanomagnets with strong shape anisotropy that mimics natural 3D spin ice materials~\cite{harris1997geometrical,ramirez1999zero,castelnovo2008magnetic}. Strong shape anisotropy makes the magnetic moments in these lithographically defined systems analogous to Ising spins. An intriguing aspect in ASI, which is attracting much interest, is the study of controlled defects and tunability of microstates in these systems. 
The impact of defects or disorder via modifying the lattice constant, shape of nanomagnets in such ASI systems remains a field of intense research~\cite{budrikis2011diversity,montoncello2018mutual,chopdekar2013controlling,drisko2017topological,keswani2018magnetization,ostman2018interaction,dion2019tunable,keswani2019micromagnetic}. Investigating \textit{individual} vertices allows detailed understanding of magnetization reversal~\cite{pohlit2016magnetic,keswani2020complex}. As a natural extension of such ASI systems, creation of novel complex geometries such as sakthi~\cite{gilbert2014emergent,lao2018classical}, tetris~\cite{gilbert2016emergent}, etc. made it possible to study frustration in different kind of geometries. As such structures can be relatively easily created, therefore, a detailed understanding of the interplay of defects and dipolar interactions may be helpful to create newer designer materials. 
 The square lattice of ASI can be considered as composed of two orthogonal sublattices of identical nanomagnets owing to their easy axes aligned along the [10] and [01] directions. Thus, square ASI geometry has fourfold symmetry and hence a deformation in the form of misalignment of the easy axis of one of the vertex nanoislands breaks the rotational symmetry of the system. As observed from simulations earlier~\cite{keswani2019micromagnetic}, this may lead to a rich and nontrivial magnetization reversal due to external magnetic field. Due to the broken symmetry engineered by introducing misalignment, the system becomes energetically inequivalent under rotation in an applied magnetic field. Thus, it may be possible to access different energy landscapes by rotating the sample with respect to an external magnetic field.  The magnetization reversal and effective anisotropy can be modified, allowing access to different microstates.\\
In order to achieve a detailed understand of the interplay of defect and dipolar interaction, in this work, we investigated the magnetization reversal for \textit{individual} defective vertex where the defect is in the form of misalignment which is artificially created. Varying the angle of the applied field with respect to the geometrical axes of different sublattices offers a route to  the generation of predictable microstates. 
In order to extract quantitative information about the energetics of the defective system under rotation and understand the magnetization reversal behavior in depth, we also performed micromagnetic simulation. Fig.\,\ref{sem}(a) shows the schematics of different types of microstates in square geometry in increasing energy (E$_{type-I}$ <...... < E$_{type-IV}$). Under the dumbbell model, each individual macrospin can be represented by a positive and negative magnetic charge. For the magnetic configurations shown in type-I and type-II (Fig.\,\ref{sem}(a)) there are two positive and two negative charges at the vertex and therefore, the vertex is chargeless, whereas the higher-energy configuration type-III and type-IV is associated with a net charge at the vertex. Fig.\,\ref{sem}(b) shows the schematics of different edge loops viz. onion (i), horse-shoe (ii) and microvortex (iii) state. Calculations based on the macrospin model show that the energy hierarchy of these states follows $E_{\rm{microvortex}}$ $<$ $E_{\rm{
horse-shoe}}$ $<$ $E_{\rm{onion}}$~\cite{keswani2018magnetization,keswani2019micromagnetic}. 
	
\section{Methods}
 The sample used for this study is an ASI vertex with closed edges. A defect in the form of misaligned island at the vertex is created which breaks the rotational symmetry. The details are shown in the schematic diagram in Fig.\ref{sem}(c). The angle between the long axis of the misaligned island and the [10] axis is chosen as 20$^{\circ}$. For this defective square ASI, stadium shaped nanoislands of dimensions 300\,nm $\times$ 100\,nm were patterned on SiO$_{2}$/Si substrate using electron-beam lithography. Thin film of Ti(5\,nm)/Ni$_{80}$Fe$_{20}$(25\,nm)/Al(5\,nm) was deposited using e-beam deposition system. Finally, the lift-off processing was used to define the magnetic nanoislands. Ti layer of 5\,nm is used for better adhesion of the Ni$_{80}$Fe$_{20}$ film on the substrate whereas Al is used as top layer to prevent oxidation of Ni$_{80}$Fe$_{20}$. The center-to-center distance between each nanoisland is 450\,nm.  For MFM imaging in presence of magnetic field, a commercial variable field module (Asylum Research) equipped with a rotatable permanent magnet was used. 
The external magnetic field was applied in-plane along [0$\bar{1}$] direction. The angle dependent studies were performed by controllably rotating the sample counter clockwise (CCW) in the fixed field as shown in Fig.\ref{sem}(c). We employed a low moment tip (magnetic moment 3$\times$10$^{-14}$\,emu) to avoid any tip induced changes in the magnetic state of the nanostructures. The distance between the sample and the MFM tip was optimized in the range of 80-100\,nm to avoid superposition of topography on the magnetic domain images. 
For the measurements, we initially fully magnetize the system along [0$\bar{1}$] and then sweep the magnetic field in the opposite direction ([01]) to image the magnetic structure in mid-transition magnetization states. The field sweep rate of 0.004 T/min was used through out the measurements. The measurements were performed at discrete fields (see below). The single domain character of the nanoislands was confirmed from MFM images where each island appears as a dumbbell of bright and dark contrasts corresponding to the head and tail of a macrospin characterizing these nanoisland. The experiments were inspired by our previously reported micromagnetic simulation work~\cite{keswani2019micromagnetic}.

\begin{figure*}
\includegraphics[width=1\textwidth]{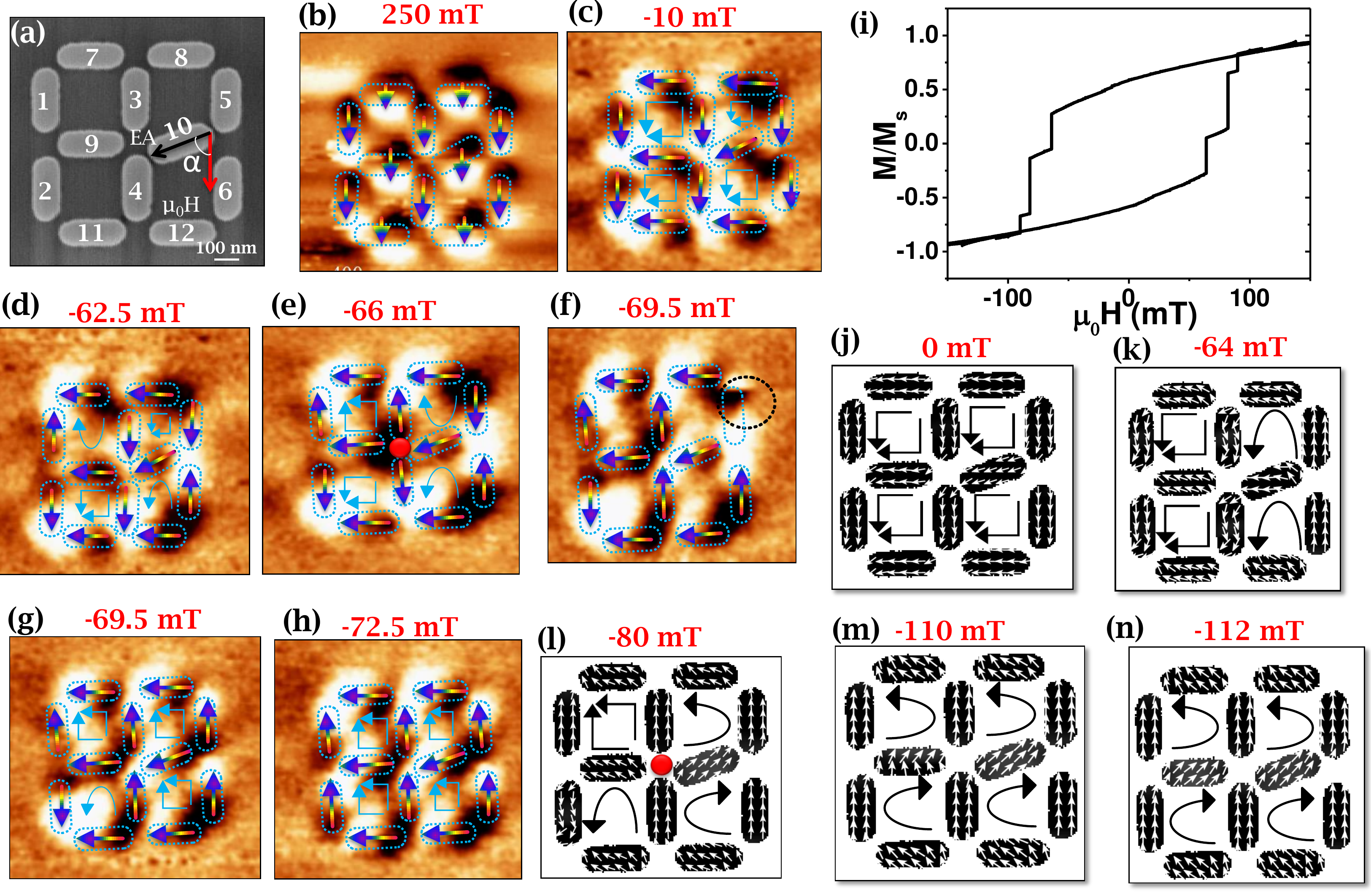}
\captionof{figure}{\label{90deg}(a) SEM image of the fabricated structure at rotation angle $\theta$ $\sim$ $0^\circ$. MFM images taken at (b) saturation, (c) near remanence and (d)-(h) intermediate fields after corresponding magnetic switchings. The arrows in the magnetic images show the orientation of magnetizations in the nanomagnets at different fields and are for guide to the eyes. (i) Simulated hysteresis loop showing 3 switchings. Micromagnetic state (j) at remanence and (k)-(n) at intermediate field.}
\end{figure*}

\begin{figure*}
	\includegraphics[width=1\textwidth]{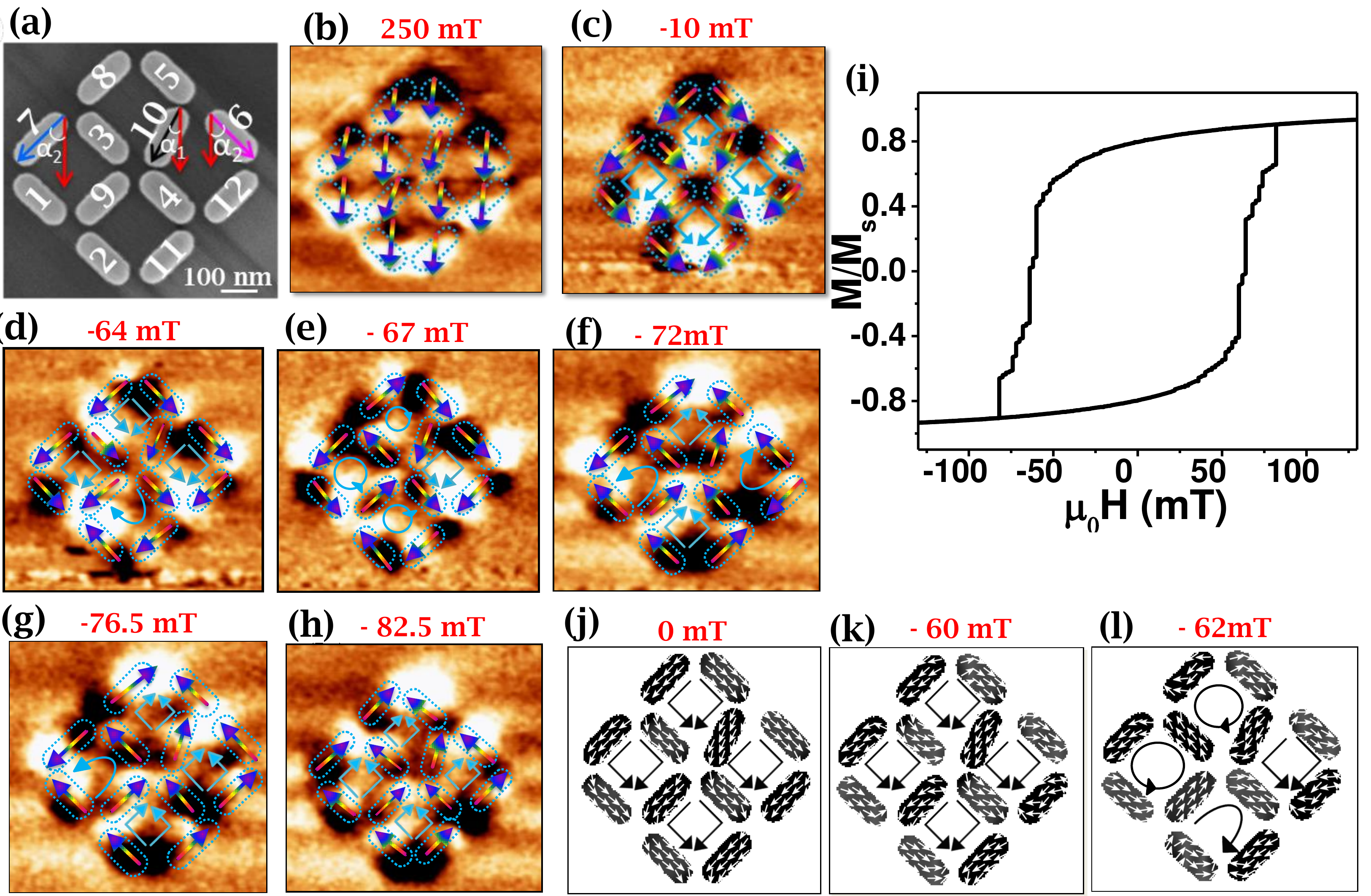}
		\captionsetup{justification=justified}
	\caption{\label{45deg}(a) SEM image of the fabricated structure for rotation angle $\theta$ $\sim$ $45^\circ$. MFM images at (b) saturation, (c) near remanence and (d)-(h) intermediate fields after corresponding magnetic switchings. (i) Simulated hysteresis loop showing multiple switchings. Micromagnetic state (j) at remanence and (k)-(l) at intermediate field. }
\end{figure*}
 Micromagnetic simulations were performed using finite difference based Object Oriented Micro Magnetic Framework (OOMMF)~\cite{donahue1999national} software at $T= $0\,K. For the simulations, the exact structures of the fabricated samples as obtained from scanning electron microscopy (SEM) images were used. The nanoislands have strong shape anisotropy ($K_{sh} \sim 7 \times 10^4$ Jm$^{-3}$) and hence we neglect the magnetocrystalline anisotropy of Ni$_{80}$Fe$_{20}$. The exchange stiffness was taken to be $A= $13\,pJ/m~\cite{principles}, and the saturation magnetization was taken to be  $M_s= 8.6 \times10^5$\,A/m, (as estimated from our SQUID measurements). The magnetization reversal was studied while sweeping the field between $\pm$200\,mT at $T= 0$\,K. To understand the switching behavior of the interacting nanomagnets and compare with our experimental data, we calculated the exact micromagnetic states of the system at every 2\,mT during the reversal.

\section{Results}
First we discuss the magnetic field dependent MFM data for the sample with field configuration as shown in the schematics in Fig.\,\ref{sem}(c).
Fig.\,\ref{90deg}(a) shows the SEM image of the fabricated structure consisting of 12 nanoislands with a misaligned island at the vertex (marked from 1 to 12). The field is applied along [0$\bar{1}$] direction. The system was initialized with the magnetization saturated parallel to an external field of 250\,mT as shown in Fig.\,\ref{90deg}(a). The MFM data at the saturated state is shown in Fig.\,\ref{90deg}(b) where the arrows indicate the direction of magnetization of the nanoislands. During reversal of the external field, the static equilibrium configuration at remanence evolves into a two-in/two-out magnetic state at the vertex with four onion states in the loops as shown in Fig.\,\ref{90deg}(c). The MFM data for $-$10\,mT $\leq$ $\mu_0 H$ $\leq$ $-$60\,mT do not show any significant change in the magnetic state of the nanoisland indicating no magnetization switching has taken place in the field range. As shown in Fig.\,\ref{90deg}(d), the first magnetic reversal is observed at $\mu_0 H$ = $-$62.5\,mT at which two diagonal islands (marked as 1 and 6) switch by forming horse-shoe states in the respective loops which has lower energy than the onion state.  
The switchings described above indicates an indirect coupling between the nanoislands 1 and 6. Similar indirect couplings of nanoisland and simultaneous switchings at a certain field were also observed in earlier reported micromagnetic simulations~\cite{keswani2018magnetization}. Next swtiching is observed at $\mu_0 H$= $-$66\,mT where the nanoisland marked as 3 switches. Due to this switching, the two-in/two-out state at the vertex changes to one-in/three-out state (see Fig.\,\ref{90deg}(e)). According to the dumbbell model, proposed by Castelnovo \textit{et al.},~\cite{castelnovo2008magnetic} a magnetic dipole can be assumed to represent magnetic charges of $+Q_{\rm{m}}$ and $-Q_{\rm{m}}$.  Thus, this reversal leads to an ice-rule-violating defect with a formation of net charge $\sum Q $ = $-2Q_{\rm{m}}$ at the vertex. At this configuration, the easy axis of misaligned island and field direction are oriented at angle $ \alpha \sim$$70^\circ$ as shown in Fig.\,\ref{90deg}(a). A large component of applied field is orthogonal to misaligned nanoisland's easy axis and thus misalinged island (island 10) doesnot switch at lower fields. On the other hand, the increased magnetostatic coupling between misalinged island (marked as 10) and island 4 leads to increased coercivity of island 4. Thus, island 3 switches before island 4 leading to generation of type-III state at the vertex. 
This charged vertex is considered as an emergent monopole state in artificial spin ice systems~\cite{phatak2011nanoscale,pollard2012propagation}. To investigate the stability of this charged state, the MFM measurement were carried out at $\sim$ 0.5\,mT interval. The charged state gets annihilated at an external field of $\mu_0 H$ = $-$69.5\,mT (see Fig.\,\ref{90deg}(f)), at which vertex island marked as 4 switches. During the same scan, the local stray field of magnetic tip induces abrupt magnetization reversal of edge island 5 as shown by dotted circle in Fig.\,\ref{90deg}(f), thereby forming three onion states and one horse-shoe state with a type-II state at vertex (see Fig.\,\ref{90deg}(g)). With the last switching of nanomagnet 2 at $\mu_0 H$ = $-$72.5\,mT, onion states are generated again in all the loops as observed at remanence which is shown in Fig.\,\ref{90deg}(h).\\
 Micromagnetic simulations of the structure show that the remanent state of the vertex is of two-in/two-out type-II spin ice state as shown in Fig.\,\ref{90deg}(j). The hysteresis in magnetization  observed from simulations performed at every 2\,mT field shows 3 sharp jumps (Fig.\,\ref{90deg}(i)).
The first jump corresponds to the switching of the islands 5 and 6 which occurs at $\mu_0 H$ = $-$64\,mT as shown in Fig.\,\ref{90deg}(k). Note here that due to the simulations carried out at every 2\,mT, an uncertainty of 2\,mT in the switching fields is to be considered. At  $\mu_0 H$ = $-$80\,mT, islands 1, 3, 4 and 10 appear to switch simultaneously (see  Fig.\,\ref{90deg}(l)). Due to these switchings, a charged vertex of type-III state (one-in/three-out) with $\sum Q = -2Q_{\rm{m}}$ is stabilzed at this field.  
  The third jump corresponds to switching of island 2 at $\mu_0 H$ = $-$90\,mT (not shown). As the reverse field is further increased, the magnetization of island 9 for which the external field is along hard axis, changes it orientation thereby turning the vertex in to type-II state at  $\mu_0 H$ = $-$112\,mT (see Fig.\,\ref{90deg}(m) and (n)). 
We note here that during reversal process, we observe bending of average local magnetization at the edges of magnetic nanoislands which may have effect in magnetic interaction between the nanoislands. The observed switching behavior is reproducible while sweeping the field in the opposite direction. 
Thus, the micromagnetic simulations reproduces the field-dependent different magnetic states of the vertex as observed in our MFM images (see Table\,\ref{Table}). However, the exact switching patterns differ in experiments and simulations which we speculate as due to changes in the local magnetization such as bending due to tip-induced effect. This may result in to a different magnetostatic coupling between the experimental nanoislands than observed in simulations. Additional effects due to minor difference in the angle $\theta$ between experiment and simulations can not be ruled out.  

In order to investigate how the applied field direction influences the interisland magnetostatic coupling and resulting magnetization reversal of the system, we changed $\theta$. 
The sample is rotated CCW by $\theta\sim$ $45^\circ$ as shown in the schematics in Fig.\,\ref{sem}(d). Profound changes are observed in reversal mechanism resulting in new microstates which were not observed earlier. At this configuration, the defective island's anisotropy axis and external applied field are oriented at an angle $\alpha _1$ $\sim$ $25^\circ$ (see Fig.\,3(a)). Thus, a large component of external field is directed along the easy axis of the misalinged nanoisland. The other nanoislands are oriented at $\alpha_2$$\sim$ $45^\circ$ with respect to the field direction (see Fig.\,\ref{45deg}). The system was saturated parallel to an external field of 250\,mT. The MFM image recorded at this field is shown in Fig.\,\ref{45deg}(b). 
As the field is reversed, the magnetization of each islands relaxes along their easy axis with a type-II configuration at the remanence. A near remanence image at $-$10\,mT is shown in Fig.\,\ref{45deg}(c). For the field less than $\mu_0 H$ = $-$63\,mT, no switching was observed. As the field approaches $\mu_0 H$ = $-$64\,mT, the nanoisland 2 switches as shown in Fig.\,\ref{45deg}(d). This converts an onion-type loop to a horse-shoe type loop. Investigating magnetic state at every $\sim$1mT, the next switching is observed at $\mu_0 H$ = $-$67\,mT, at which islands 3, 8 and 9 switch simultaneously (see Fig.\,\ref{45deg}(e)), thereby converting  two onion and one horse-shoe into three microvortex loops. With these switchings, the chain of nanoislands consisting of islands 2, 3, 9 and 8 all have their major component of magnetization oriented along the external field direction. Together with other islands, viz., 5, 10, 4 and 11, they form a larger loop which is energetically favorable. The microvortex state consists of sublattices with opposite magnetizations only. Thus, the three microvortices lead to reduction in the no. of head-to-head (tail-to-tail) configurations and hence are energetically more favourable. Interestingly, these switchings also lead to the creation of a type-I state at the vertex. Considering that the square ASI vertex of type-I state of lowest possible energy was never observed during magnetization reversal for the case of $\theta=0^\circ$, this is a remarkable observation. This demonstrates a possible way of achieving the ground state of an ASI-vertex. As the field is further increased to $\mu_0 H$ = $-$72,mT, six islands (1, 4, 5, 6, 10 and 11) switch together and eventually type-II state is generated again at the vertex as illustrated in Fig.\,\ref{45deg}(f). It is clearly evident that the rotation of the system with respect to the field angle increases the component of applied field along the easy axis of different nanoislands resulting in multiple switching and coupling of different nanoislands. The next switching occurs at $\mu_0 H$ = $-$76.5\,mT and $-$82.5\,mT, at which island 12 (see Fig.\,\ref{45deg}(g)) and island 7 (see Fig.\,\ref{45deg}(h)) switch respectively. The corresponding simulated hysteresis loop, as shown in Fig.\,\ref{45deg}(i), depicts the gradual rotation of magnetization of the nanoislands followed by six sharp jumps. We show only 3 micromagnetic states here (Fig.\,\ref{45deg}(j)-(l)). Fig.\,\ref{45deg}(j) illustrates the micromagnetic state at remanence with four onion states with a type-II state at the vertex. As the field is increased in the reverse direction, again in this case, we observe the edges of the stadium shaped nanomagnets exhibiting curling of the local magnetization directions (shown in  Fig.\,\ref{45deg}(k)). Fig.\,\ref{45deg}(l) shows  micromagnetic state captured just after the first switching of island 3 which takes place at $\mu_0 H$ = $-$62\,mT. The curl or bending of local magnetization at the edges of some nanoislands are clearly observed here. In addition to these details of local magnetization within the individual nanoislands, importantly, we observe that the magnetization switching of the island 3 converts the vertex of type-II to a type-I.  
 Here again, we find that the exact sequence of the switching of the nanoislands doesnot match with our experiments however the simulation results are consistent with the experimental observation of the evolution of the vertex state as a function of applied field.    
    \begin{figure*}
  	\includegraphics[width=0.8\textwidth]{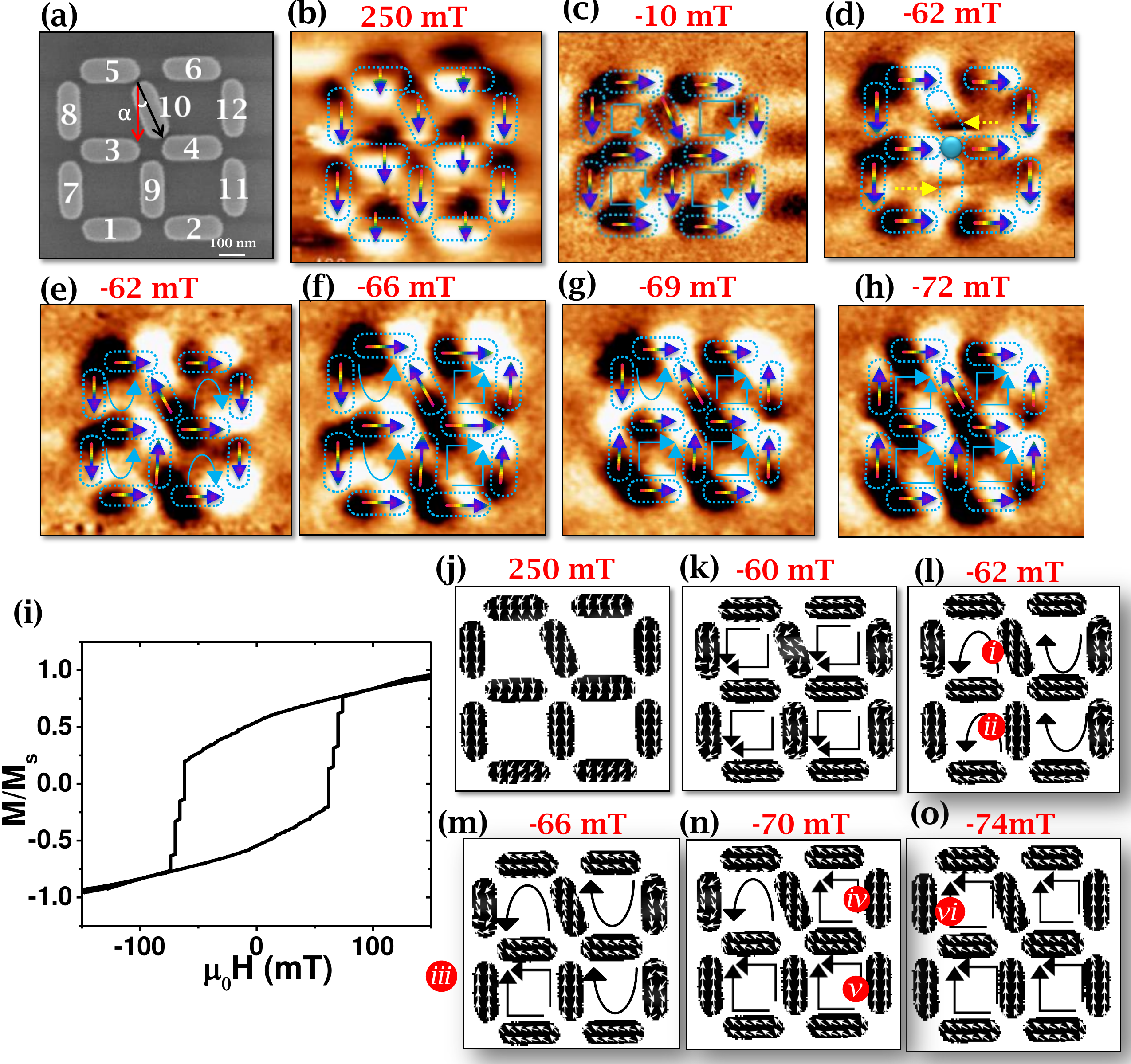}
  	\captionsetup{justification=justified}
  	\caption{\label{mfm1}(a) SEM image of the fabricated structure at rotation angle $\theta$ $\sim$ $90^\circ$. MFM images taken at (b) saturation, (c) near remanence, (d)-(h) intermediate fields after corresponding magnetic switchings. Two tip-induced switchings are indicated by dotted arrows in (d). (i) Simulated hysteresis loop showing four switchings. Corresponding micromagnetic state at (j) saturation and (k)-(o) at intermediate field.}
  \end{figure*}
Thus, our experimental and simulation results for this relative field configuration suggests an increased stability of the system at this configuration resulting in to the creation of the lower energy type-I state at the vertex. The initial and final reversal of magnetization is observed at a  greater field value ($\mu_0 H$ = $-$64\,mT and $\mu_0 H$ = $-$82.5\,mT) for this orientation of $\theta$ $\sim$ $45^\circ$. The data show that the microstates of the system are strongly influenced by the geometric arrangement of the islands with respect to the applied field. \\

We next discuss our results for rotation angle $\theta$ $\sim 90^\circ$. The same measurement field protocol was followed as discussed in the above 2 cases. At this configuration, the external field makes an angle $\alpha$$\sim20^\circ$ with the easy axis of misalinged nanoisland, while other nanoislands are oriented along or perpendicular to the applied field (see Fig.\,\ref{mfm1}(a)). 
An MFM image of the system at saturation ($\mu_0 H$ = $250$\,mT) is shown in Fig.\,\ref{mfm1}(b). In this case also, we observe that the vertex at the remanence is of type-II. This is also observed at $\mu_0 H$ = $-$10\,mT (Fig.\,\ref{mfm1}(c)) where no switching has taken place so far. An interesting switching behavior is observed at $\mu_0 H$ = $-$62\,mT. While scanning at this field, we observe that the local stray field of the MFM tip induces a switching of magnetization of island 9 thereby creating a three-in/one-out type-III state at the vertex (see Fig.\,\ref{mfm1}(d)). This tip induced switching during scanning is observed as white-white patches at the ends of island 9 which otherwise exhibits black-white patches as can be seen in Fig.\,\ref{mfm1}(c). Similar color contrasts at both ends indicate that the switching occured as the scanning progressed over the single-domain nanoisland. Interestingly, as the scanning progresses further upwards, another tip induced switching takes place for the island 10. This switching converts the vertex from type-III to type-II which is chargeless. 
The two tip-induced switchings are indicated by yellow arrows in Fig.\,\ref{mfm1}(d). Thus, these abrupt switchings convert a type-III state (charged state) to type-II (chargeless state) during the same scan (see Fig.\,\ref{mfm1}(e)). It is important to note here that the same magnetic tip as well as scan height were used for the cases, viz., for $\theta = 0^\circ$ or $45^\circ$ as discussed above. But no such tip-induced switching was observed for those two cases. 
This reversal changes all the onion states to lower energy horse-shoe states as shown in Fig.\,\ref{mfm1}(e). The second switch is observed at $\mu_0 H$ = $-$66\,mT  at which islands 11 and 12 switch simultaneously as shown in Fig.\,\ref{mfm1}(f). At $\mu_0 H$ = $-$69\,mT, island 7 switches thereby forming three onion states and one horse-shoe state (see Fig.\,\ref{mfm1}(g)). The fourth switch corresponds to switching of island 8 at $\mu_0 H$ = $-$72\,mT which restores onion states again in all the loops (see Fig.\,\ref{mfm1}(h)). \\
Further insights in to the interesting switching behavior is gained by performing micromagnetic simulations for this field configuration given by $\theta=90^{\circ}$. The simulated hysteresis loop shows that the magnetization reversal of the system in this case takes place via four sharp jumps indicative of four switchings (see Fig.\,\ref{mfm1}(i)). A closer look at the micromagnetic states near these switching fields shows that magnetization of island 10 orient in a zig zag way at $-$60\,mT, most likely due to a strong competition between the Zeeman and the anisotropy energy at this configuration. It is plausible to assume that the energy barrier is reduced as a result of for the metastable charged state as observed experimentally in Fig.\,\ref{mfm1}(d), is due to islands 10  with curled magnetization (see Fig.\,\ref{mfm1}(k)). It appears that the curled magnetization of island 10 reduces the energy barrier thereby triggering a switching by the stray field emanating from the MFM tip.  
Similar to the experimental observations, simulation results also show that the switchings of island 9 and 10 results in to the conversion of the respective onion type loops to lower-energy horse-shoe type loops.
 Figs.\,\ref{mfm1}(l)-(o) shows micromagnetic states at four different switching fields. It is clear the simulation results exhibit the same no. of horse-shoe as well as onion type loops formed after each switching.  Moreover, the sequence of the switching of different nanoislands as observed in simulations follows the experimental results. Thus, we find that the results of the micromagnetic investigations for this orientation are in excellent agreement with switching observed experimentally. Different vertex states formed at respective fields are mentioned in Table\,\ref{Table}. In order to investigate systematically different states generated as a function of $\theta$, we performed simulations for five other $\theta$ values, viz., 30$^{\circ}$, 60$^{\circ}$, 135$^{\circ}$, 160$^{\circ}$ and 180$^{\circ}$, respectively.
  \begin{table}
  	\begin{tabular}{ |p{1.6cm}|p{3cm}|p{3.8cm}|}	
  		\hline
  	\textbf{Angle of rotation ($\theta$)} &  \multicolumn{2}{c|}{\textbf{Magnetic states at intermediate field}}\\
  	\cline{2-3}
  	&\textbf{Experimental}&\textbf{Simulation}\\
  		\hline
  		$0^{\circ}$   & type-II$^\ast$ and type-III (66\,mT$-$69.5\,mT) & type-II$^\ast$ and type-III (82\,mT$-$100\,mT) \\
  		\hline
  		$30^{\circ}$  & - & type-II$^{\ast}$\\
  		\hline
  		$45^{\circ}$& type-I (67\,mT$-$72\,mT) and type-II$^\ast$ & type-I (62\,mT$-$66\,mT) and type-II$^\ast$\\
  		\hline
  	    $60^{\circ}$ &- &type-I (62\,mT$-$66\,mT) and type-II$^\ast$\\
  		\hline
  		$90^{\circ}$&  type-II$^{\ast}$ &  type-II$^{\ast}$\\
  		\hline
  		$135^{\circ}$&-& type-II$^{\ast}$\\
  		\hline 
  		$160^{\circ}$  & - & type-III$^\ast$ and  and type-II (74\,mT-142\,mT)\\
  		\hline
  		$180^{\circ}$& - & type-II$^\ast$, type-III (100\,mT-200\,mT) and vortex states ( 18\,mT to $-$122\,mT)\\
  		\hline
  		
  	\end{tabular}

  	\caption{\label{Table}Table summarizes the magnetic states at intermediate fields corresponding to different rotation angles. ($^\ast$ refers to vertex state observed for all other  fields.)}

  \end{table}
For all eight rotation angles studied here, the vertex always evolves to type-II state at remanence as well as after complete magnetization reversals. However, our observations suggest that through the rotation of the system with respect to the external field, it is possible to probe a rich energy landscape with different microstates for this defective vertex system. The corresponding magnetic states of the vertex for various rotation angles studied in this work are summarized in Table\,\ref{Table}.\\

\begin{figure*}
		\includegraphics[width=1\textwidth]{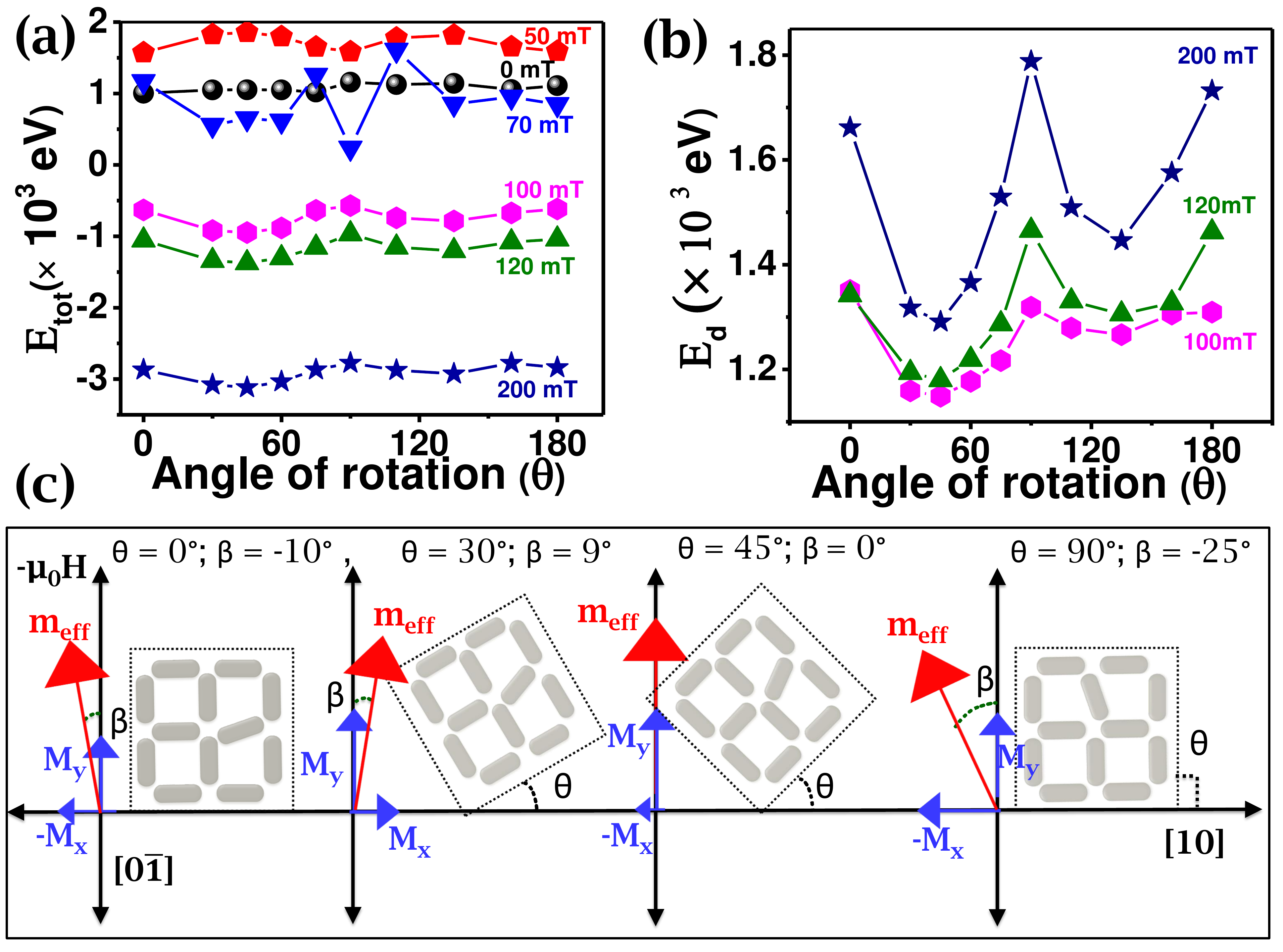}
		\caption{\label{dipolar} Variation of (a) total energy ($E_{\rm{tot}}$) and (b) dipolar energy ($E_d$) with angle of rotation($\theta$) at different external field. (c) Illustration of variation of $\beta$ with angle of rotation($\theta$) at $\mu_0 H$ = $-$100\,mT.} 
\end{figure*}

Next, we discuss the energetics of the system as it evolves with the variation of $\theta$. Fig.\,\ref{dipolar}(a) shows the total energy $E_{\rm{tot}}$ of the system as a function of $\theta$ for six different values of the external field, viz., $\mu_0 H$ = 0\,mT ($E_{0}$),  $\mu_0 H$ = 50\,mT ($E_{50}$), $\mu_0 H$ = 70\,mT ($E_{70}$), $\mu_0 H$ = 100\,mT ($E_{100}$), $\mu_0 H$ = 120\,mT ($E_{120}$) and $\mu_0 H$ = 200\,mT ($E_{\rm{sat}}$) respectively. $E_{0}$ is the energy of the system at remanence which expectedly, remains almost constant for all $\theta$. At higher fields altered dipolar interactions among the nanomagnets leads to variations in the net energy as $\theta$ changes. This behavior of energy remains qualitatively similar for all fields which are less than the switching fields. Note that the switching fields for all the configuration as observed from simulations are in the range of 60\,mT - 85\,mT (see Table\,\ref{Table}). It is evident from Fig.\,\ref{dipolar}(a) that as the field increases beyond the switching field, the shape of the energy profile manifests itself in a very clear form. For clarity, only three energy profiles in this higher field range, viz., $E_{100}$, $E_{120}$  and $E_{\rm{sat}}$ are plotted. At these high fields, the energy becomes negative for all values of $\theta$. As shown in Fig.\,\ref{dipolar}(a), the profiles at these fields show that the rotation of the system leads to an asymmetric energy landscape with a global minima refering to the lowest energy point on the entire energy landscape at $\theta \sim45^{\circ}$ and a local minima at $\theta \sim135^{\circ}$. 

The energy profiles for fields within the switching field regime indicate multiple features reflecting the changes in energy due to the switchings of nanomagnets in this field regime. A typical example of such an energy profile in this field regime is shown by plotting $E_{70}(\theta)$. 

Such variations in energy can be understood by invoking Stoner-Wohlfarth model for the dipolar coupled system for which we consider effective anisotropy ($K_{\rm{eff}}$), effective magnetic moment ($m_{\rm{eff}}$) etc. of the system consisting of 12 similar nanoislands. 
In that case, the total energy ($E_{\rm{tot}}$) of the system in presence of field is given as sum of anisotropy energy ($E_a$), dipolar interaction energy ($E_d$), and Zeeman energy ($E_{z}$)  of the system, i.e., 

\begin{equation}
E_{\rm{tot}} = E_a + E_d +E_z
\label{sw}
\end{equation}

where, $E_z\,=\,\mu_0 m_{eff}H cos\beta$, 
$H$ the external field and $\beta$ is the angle between effective magnetic moment $m_{\rm{eff}}$ and applied field $H$. The first term  in eqn.\ref{sw} remains constant as $\theta$ is varied. On the other hand, the second term, $E_d$ shows clear dependence on $\theta$ in presence of field as depicted in Fig.\,\ref{dipolar}(b) which shows the variations of $E_d$ for $\mu_0H$\,=\,100\,mT, 120\,mT and 200\,mT, respectively. It is evident that corresponding dipolar energies also show global minima at $\theta$ = $45^\circ$ at these fields.  
The third term in eqn. (1) depends on $\beta$ which dominates at higher fields and in turn depends on $\theta$ as discussed below. 


 To understand the energy profile at magnetic fields higher than the switching fields, we evaluate $\beta$ for varying $\theta$ for our system. From the magnetization of individual nanoislands in the coupled system for a given $\theta$ and $\mu_0 H$ as obtained from the simulations, we determine the corresponding direction of $\vec{m_{\rm{eff}}}$($\theta$) for different fields applied in the direction of [01]. 
 Fig.\,\ref{dipolar}(c) shows the orientation of $m_{\rm{eff}}$ for four different $\theta$ values calculated for $\mu_0 H = -100$\,mT. The results for the three fields are tabulated in Table\,\ref{2}.

\begin{table}
\begin{tabular}{ |c|c|c|c|c|c| } 

		\hline
		$\theta$ & $\beta$($-$100 mT)& $\beta$($-$120 mT) & $\beta$($-$200 mT)\\
		\hline	 
		0$^\circ$ & $-10^\circ$ & $-5^\circ$ & 10$^\circ$\\
		\hline		
		30$^\circ$	& $9^\circ$ & $8^\circ$ & $6^\circ$\\
		\hline
		$45^\circ$ & 0$^\circ$ & 0$^\circ$ & 0$^\circ$\\
		\hline
		$60^\circ$ & $-9^\circ$ & 0$^\circ$ & $-4^\circ$\\
		\hline
		$90^{\circ}$ & $-25^\circ$ & $-17^\circ$ &  10$^\circ$\\
		\hline
	    $135^{\circ}$ & $0^\circ$ & $-2^\circ$ &  $0^\circ$\\
		\hline
		$160^{\circ}$ & $-14^\circ$ & $-11^\circ$ & $-2^\circ$\\
		\hline
		$180^{\circ}$ & $-16^\circ$ & $-12^\circ$&  $0^\circ$\\
		\hline
	\end{tabular}
	\caption{\label{2}Table summarizes values of angle $\beta$ at applied field $\mu_0 H$ = $-$100\,mT, $-$120\,mT and $-$200\,mT respectively, for different rotation angles.}
\end{table}

We find that for $\theta$ = $45^\circ$, $m_{\rm{eff}}$ orients along the direction of the applied field (i.e., $\beta=0^{\circ}$) which is the condition for the minimum energy state. Manifested of this is observed as a global minimum in the profiles of $E_d$ and in turn $E_{\rm{tot}}$. 
In the micromagnetic behavior, interestingly, we observe the vertex to acquire the ground state (type-I) configuration at this angle of rotation which most likely is related to the global minimum of the energy. We note further that the micro-vortex type edge loops which have the lowest energy are also observed for this angle. 
Thus, these results clearly demostrate that the angle between external field and the geometric axis of an ASI vertex can be considered as an important parameter to explore the energetics of the system.  


\section{Conclusions}
Our results show that the ASI vertex structure with broken rotational symmetry presents an interesting system for experimental exploration. The broken symmetry engineered by introducing misalignment leads to an energetically inequivalent system under rotation in an applied magnetic field and allows easy access to different energy landscapes. By rotating the samples in an applied field, we are able to stabilize different vertex configurations, from magnetically chargeless type-I to charged type-III states. 
Our results suggest the role of intricate interplay of defect and dipolar interactions in predictably stabilizing different vertex states which may be of interest to explore electronic transport behavior in such clearly defined vertices. 

\begin{acknowledgments}
	 We gratefully acknowledge the technical support of Masahide Tokuda for the deposition of Ni$_{80}$Fe$_{20}$ film. N.K. wishes to thank Manish Anand for fruitful discussions. N.K. is also thankful to University Grant Commission (UGC) Govt. of India, for providing research fellowship. P.D. acknowledges the partial finanical support through collaborative research and education under IIT Delhi-BNERC, Toyo University's joint Bio-Nano Mission program. Part of the work was carried out at the Nano Research Facility (NRF) and High Performance Computing (HPC) Centre of IIT Delhi. 
\end{acknowledgments}
\bibliography{bibitem}
	
\end{document}